\documentclass[aps,showpacs,twocolumn]{revtex4}
\usepackage{graphics}
\usepackage{amsmath}
\usepackage{amsfonts}
\usepackage{pifont}
\usepackage{mathrsfs}

\begin{document}
\newcommand{\bstfile}{aps} 
\newcommand{\bibs}{IntTheoryRefs,BibFile}

\title{A Self Healing Model Based on Polymer-Mediated Chromophore Correlations}
\author{Shiva K. Ramini and Mark G. Kuzyk}
\address{Department of Physics and Astronomy, Washington State University \\ Pullman,
Washington  99164-2814}
\date{\today}

\begin{abstract}
Here we present a model of self healing in which correlations between chromophores, as mediated by the polymer, are key to the recovery process.  Our model determines the size distribution of the correlation volume using a grand canonical ensemble through a free energy advantage parameter.   Choosing a healing rate that is proportional to the number of undamaged molecules in a correlated region, and a decay rate proportional to the intensity normalized to the correlation volume, the ensemble average is shown to correctly predict decay and recovery of the population of disperse orange 11-DO11 (1-amino-2-methylanthraquinone) molecules doped in PMMA polymer as a function of time and concentration as measured with amplified spontaneous emission and linear absorption spectroscopy using only three parameters that apply to the full set of data. Our model also predicts the temperature dependence of the process.  One set of parameters should be characteristic of a particular polymer and dopant chromophore combination.  Thus, use of the model in determining these parameters for various materials systems should provide the data needed to test fundamental models of the underlying mechanism responsible for self healing.

\end{abstract}


\maketitle

\vspace{1em}

\section{Introduction}

Photodegradation is an inherent problem of materials that are used for high-light-intensity applications such as organic lasing media\cite{dyuma92.01,popov.98.01} or materials for use in nonlinear-optical devices.\cite{zhang98.01,galvan20.01} Peng first observed the recovery of fluorescence after photodegradation of a dye-doped polymer optical fiber.\cite{Peng98.01} Howell observed full recovery after photodegradation of Amplified Spontaneous Emission (ASE) from Disperse Orange 11 (DO11) dye doped in PMMA polymer.\cite{howel01.01,howel02.01} The DO11 molecule was shown not to recover in liquid solution.\cite{howel04.01} An intriguing demonstration that laser cycling could make materials more efficient and robust to future degradation\cite{zhu07.01,kuzyk07.02} motivated further research.

The mechanism responsible for the decay and recovery process is still not well understood.\cite{embay08.01} The purpose of this contribution is to develop a model based on physically realistic assumptions that take into account all observations.  The model that we propose correctly predicts all observations, makes new predictions that can be tested, and is expressed in terms of only three parameters that may be calculable from first principles.

\section{Background}

ASE is the most sensitive probe of photodegradation and self healing that has been extensively used to characterize PMMA polymer doped with DO11 chromophores.\cite{embay08.01}   While not as well studied, decay and recovery of the AF455 chromophore has been observed using two-photon absorption (TPA) spectroscopy.\cite{zhu07.01,kuzyk07.02}  Since ASE and two photon absorption are nonlinear-optical processes, this makes them sensitive probes but can also lead to larger experimental uncertainties.  Additionally, identically-prepared polymers tend to vary from sample to sample, so, it may be difficult to reproduce data runs with a high degree of precision.

\begin{figure}
\includegraphics{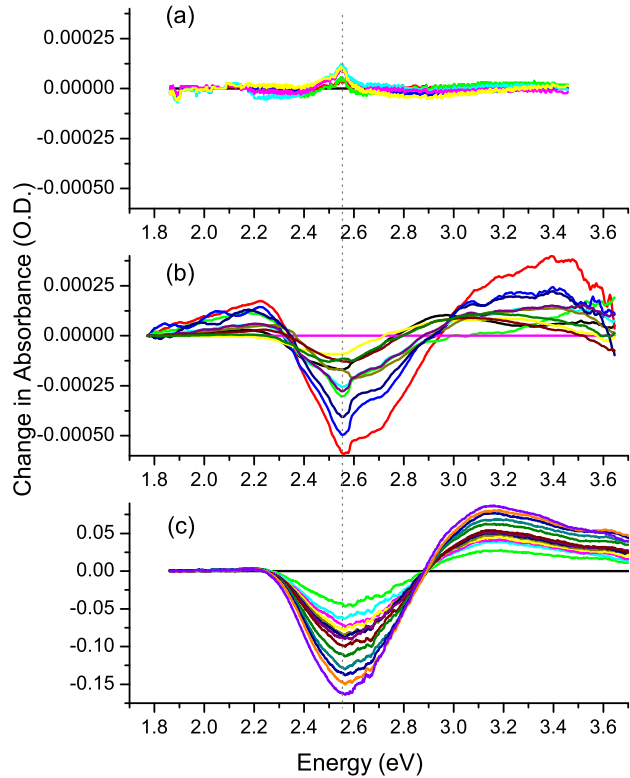}
\caption{(a) Change in absorbance as a function of dye concentration with respect to the lowest DO11 dye concentration of DO11 dissolved in the liquid monomer MMA. (b) Same as (a) but for DO11 doped in PMMA polymer. (c) Change in the absorbance due to photodegradation in DO11/PMMA.}
\label{fig:abschange}
\end{figure}

Keeping these difficulties in mind, it is nevertheless possible to determine common features over the full set of observations that hint at the mechanisms responsible. In particular, the ASE studies suggest the following,
\begin{enumerate}
\item{Dye molecules that irreversibly photodegrade in a liquid are found to self heal after photodegradation in a polymer.\cite{howel02.01,howel04.01}}
\item{Full recovery of the ASE signal after self healing is observed only for dye concentrations near the saturation limit in the polymer.  The degree of recovery decreases at lower concentrations.\cite{embay08.01}\label{item:SaturationRecovery}}
\item{For a particular concentration, the decay rate depends more or less linearly on the pump intensity but the recovery rate is a constant.\cite{embay08.01}}
\item{The two-photon absorption cross-section of AF455 decays at a rate in proportion to the intensity, but recovers at a single recovery rate provided that the sample is not severely damaged\cite{zhu07.01}.}
\item{Linear dichroism is constant during decay and recovery, suggesting that molecular reorientation is not responsible.\cite{embay08.01}}
\item{Decay of ASE signal is accompanied by a change in the absorption spectrum with an isosbestic point, suggesting that the decay product is a different species, or a perturbed version of the chromophore, and rules out the possibility that dye diffusion is responsible\cite{embay08.01} unlike in some other systems\cite{lal99.01}.}
\item{Evolution of the spatial population profile of a burn mark during recovery does not fit the diffusion model -- additional evidence that diffusion is not the cause of recovery.\cite{ramin11.01}}
\item{At high-enough intensities that result in visible damage (burn marks and laser ablation), the sample {\em appears} to decay irreversibly, but there is some evidence that the recovery times in these cases are much longer than for the regime in which the sample degrades by a small amount.\cite{ander11.01,desau09.01}}
\item{After cycles of photodegradation and recovery, the ASE intensity increases and the decay time constant increases relative to a non-cycled sample.\cite{howel04.01,zhu07.01}}
\item{As a function of temperature, the ASE intensity decreases, but, the change in linear absorption spectrum peaks at a different wavelength than when the material is photodamaged, suggesting that a new species is populated at higher temperature (as we will show below).  Furthermore, the species found at elevated temperature recovers more quickly than a photodamaged molecule.  Thus, the thermally-populated species - which we call a bystander state, is not a damaged molecule.}
\item{The decay rate is observed to decrease with increasing concentration, as we discus later (and shown in Figure \ref{fig:ConcDecayFits}); but, the recovery rate increases with concentration.\label{item:DecayRateConcentration}}
\item{The change in the linear absorbance spectrum with concentration peaks at the same wavelength as does the change in the spectrum as the material photodamages. No such change in the normalized linear absorbance spectrum is observed as a function of concentration in MMA liquid as shown in Figure \ref{fig:abschange}.  This suggests that the damaged species and dye aggregates may be related.\label{item:AbosrbanceWithDecaySameAsConcetration}}

\end{enumerate}

The key observations that drive our model are that (1) full recovery requires high chromophore concentration (Item \ref{item:SaturationRecovery} above), (2) the decay rate decrease with concentration (Item \ref{item:DecayRateConcentration} above) and (3) optical absorption spectrum evolves as a function of time in the same way that it changes with concentration (Item  \ref{item:AbosrbanceWithDecaySameAsConcetration} above).  These observations, along with the fact that the polymer plays a critical role, suggests that the interactions between chromophores as mediated by the polymer are responsible for self healing.  As such, we call our new model the {\em correlated chromophore model}, which is a generalization of the simpler {\em embedded chromophore model} used in the past.\cite{embay08.01} The role of temperature will be taken into account using the grand canonical partition function, which includes correlation scales that depend on interactions between chromophores.

\section{The Model}

In the following sections, we begin by describing the embedded chromophore model, generalize it to include correlations between chromophores, apply the grand canonical ensemble to get the temperature-dependent distribution of the size of the correlated regions, and combine these into the full theory of the decay and healing process.

\subsection{Embedded Chromophore Model}

Consider a system of $N$ {\em non-interacting molecules}, $n$ of which are undamaged.  Assuming that there is only one damaged species, the damaged population is $N-n$.  If the recovery rate is $\beta$ and the decay rate proportional to the intensity, $I$, and given by $\alpha I$, then the evolution of the population is given by,
\begin{equation}\label{NonInteractingRates}
\frac {d n } {dt} = - \alpha I n + \beta \left( N-n\right),
\end{equation}
which has a solution for $I \neq 0$ of
\begin{equation}\label{nDecay}
\frac {n(t)} {N} = \frac {\beta} {\beta + \alpha I} + \frac {\alpha I } {\beta + \alpha I} \cdot e^{- \left( \beta + \alpha I \right) t },
\end{equation}
where the sample is assumed to be originally pristine at $t=0$.

The population of undamaged molecules, starting when the pump is turned off at time $t_0$, is given by
\begin{equation}\label{nRecover}
\frac {n(t)} {N}  = 1 - \left( 1- \frac {n(t_0)} {N} \right) e^{- \beta \left( t - t_0 \right) },
\end{equation}
where $n(t_0)$ is the initial undamaged population at time $t_0$.  This model is consistent with the time evolution of the ASE intensity in DO11\cite{embay08.01} and the TPA crossection of AF455\cite{zhu07.01} chromophores at fixed concentration and temperature.  Even though the population model is by necessity dependent on more than two populations (i.e. excited electronic and vibronic states of each species must play a role), if optical excitations and de-excitations are fast compared with the photodegradation and recovery process, the dynamics of these other states can be ignored.  Note that the population of the chromophore and the decay product can be differentiated by their unique optical properties, such as ASE efficiency and TPA cross section.

\subsection{Correlated Chromophore Model}

The noninteracting model given by Equation \ref{NonInteractingRates} can be phenomenologically generalized by allowing the rates $\alpha$ and $\beta$ to depend on intensity -- for example, by expressing them as a series in the intensity if higher order contributions are small.  $\alpha$ is independent of intensity if the optical damage process is dominated by one-photon absorption.  If two-photon absorption is important, than a correction term linear in the intensity must be added to the theory.  Similarly, it is possible that the recovery rate is accelerated or suppressed in the presence of light, in which case an intensity-dependent correction term needs to be added to $\beta$.  The data for DO11/PMMA and AF455/PMMA at fixed concentration and temperature agree with the theory for constant $\alpha$ and $\beta$, so the higher-order correction terms are ignored in this paper, but can be easily generalized if needed.

The observation that decay and recovery kinetics depend on the concentration can be taken into account by making $\alpha$ and $\beta$ a function of the concentration of the chromophores and their decay products.  In particular, Figure \ref{fig:abschange} shows that the absorption spectrum changes as a function of DO11 concentration in PMMA polymer, an indication that the DO11 molecules are interacting with each other.  The fact that higher concentration samples show accelerated recovery\cite{embay08.01} and decreased decay rates suggests that interactions between molecules may be responsible for self healing.  The nature of this interaction is not important in building a phenomenological model that depends on concentration and characteristic energies.  Indeed, the phenomenological parameters that we will define in our model would be calculable from first principles as tests of the nature of these interactions.

\begin{figure}
\includegraphics{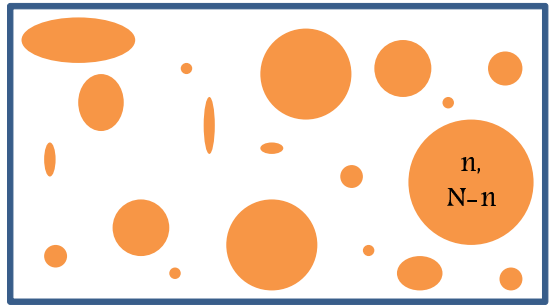}
\caption{Domains of dye in polymer.}
\label{fig:domain}
\end{figure}

In the generalized model, we redefine $N$ to be the number of molecules that are associated with each other, be it through physical aggregation of DO11 molecules into dimers or microcrystalites, a correlated region of dyes that interact through the polymer as mediated by phonons, locally oriented domains, or altogether new physics.  For the purposes of this paper, we will generically call such a correlated region a domain.  Each domain will have its own characteristic decay and recovery rate depending solely on the domain size; and, the domain size distribution will be determined by the grand canonical ensemble.  The observed bulk behavior will then be given by the ensemble average.  Figure \ref{fig:domain} shows a cartoon representation of a collection of domains with one of the domains specified with $N$ molecules, $n$ undamaged molecules and $N-n$ damaged species.

First, we focus on a single domain of fixed size $N$.  We propose that the recovery rate of a damaged species will be accelerated in the presence of undamaged molecules in proportion to the number of undamaged molecules.  Thus, in a bigger domain, the recovery rate will be larger than in a small domain.  On the other hand, a small domain with mostly undamaged molecules will recover at a faster rate than larger domains that are populated with a preponderance of degraded molecules.

There are several mechanisms that could lead to such behavior.  For example, if the undamaged molecules are strongly interacting with each other, forming a damaged species will cost energy.  The more neighbors, the higher the energy cost, leading to a slower decay rate and faster recovery rate. The nature of these interactions is not yet understood, but there are many potential candidates, from electric forces that are typically responsible for aggregation into dimers and microcrystalites to spin statistics that cause Bose-Einstein condensation or perhaps new unknown mechanisms.  Based on experimental observations, we find that the decay rate $\alpha I$, depends inversely on the number of molecules in a domain, or that the decay rate is given by $\alpha I /N$.  These two generalization of Equation \ref{NonInteractingRates} leads to
\begin{equation}\label{InteractingRates}
\frac {d n } {dt} = \beta n (N - n) - \frac{\alpha I} N n  .
\end{equation}

Integrating Equation \ref{InteractingRates}, yields
\begin{equation}\label{RecoverN}
n = \frac { \left( N - \alpha I / \beta N \right) n_0} {n_0 + \left( N - n_0 - \alpha I / \beta N  \right) \exp \left [ - \left(N \beta - \alpha I/N \right) t\right] } ,
\end{equation}
where $n_0$ is the initial undamaged population at $t=0$.  When the pump is turned off ($I=0$), Equation \ref{RecoverN} approaches $n=N$ at infinite time.  With the pump on, at the infinite time limit, the population $n$ approaches $N - \alpha I / \beta N$ provided that the intensity is below  the critical intensity $I < I_C \equiv N^2 \beta/\alpha  $ and vanishes at large time when the intensity is above this critical value.  Past work showed that for $I<I_C$, the ASE intensity for DO11 in PMMA polymer did indeed asymptotically approach a nonzero value at long times.\cite{{embay08.01}}

With $I=0$, the population recovers according to
\begin{equation}\label{InteractingRates2}
\frac {d n } {dt} = \beta n (N - n).
\end{equation}
With $n(t_0)$ as the undamaged population after the decay,
\begin{equation}\label{RecoverN2}
n = \frac { N n(t_0)}{n(t_0) + \left[ N - n(t_0)\right] e^{\left( -N \beta t\right)}} ,
\end{equation}
or,
\begin{equation}\label{RecoverN2}
n = \frac { N }{1 + \left[ \frac N {n(t_0)}- 1\right] e^{\left( -N \beta t\right)}} .
\end{equation}

\subsection{Bystander state}
\label{sec:bystander}

\begin{figure}
\includegraphics{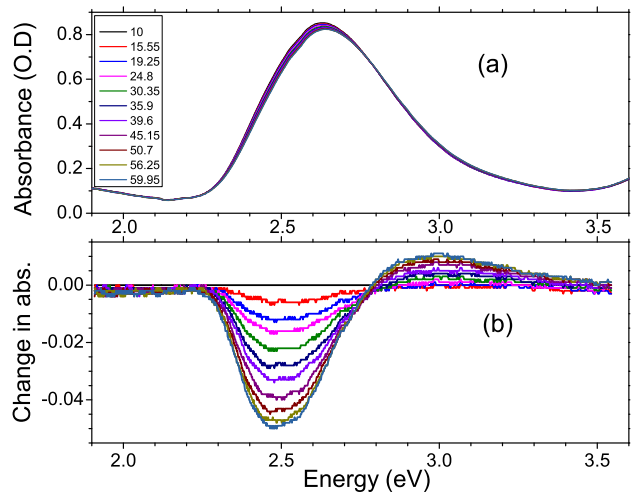}
\caption{(a) Absorbance spectrum of DO11 molecules when doped in PMMA at various temperatures. (b) Difference in absorbance with respect to absorbance at $10^0C$. As the sample temperature increases, the population of the bystander state increases as can be seen from the growing peak at 3.0eV while the population under main peak decreases.}
\label{fig:abstemp}
\end{figure}

A plot of the optical absorbance as a function of temperature, as shown in Figure \ref{fig:abstemp}a, shows the main absorption peak decreasing as a new peak grows in proportion to the main peak's decrease. The difference between each spectrum and the initial spectrum at $T=10^o C$ is shown in Figure \ref{fig:abstemp}b.  The isosbestic point between these two regions is an indication that the decreasing peak reflects a decrease in the population of one species while the growing peak is a sign of its conversion to a new species.  This process is found to be associated with a decrease in the ASE peak, an indication that the population of DO11 molecules that generate ASE are being converted to ones that do not contribute to ASE.

The positions of the two peaks in Figure \ref{fig:abstemp}b are at different energies than the corresponding peaks found during photodegradation, as shown in Figure \ref{fig:abschange}c, suggesting that the product formed at higher temperature is not necessarily the damaged state.  However, the most convincing evidence that an increase in temperature does not produce a damaged state is that the process is instantaneously reversible, that is, the spectrum change follows the temperature with negligible delay.  In contrast, recovery of the damaged species takes several hours, even at elevated temperatures.  We take this large difference in time scales as strong evidence that the bystander state is not a damaged species.

\begin{figure}
\includegraphics{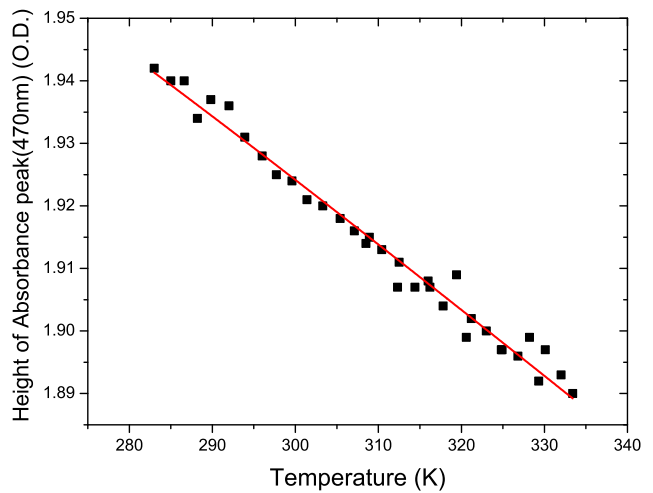}
\caption{Height of the absorbance peak is plotted as a function of temperature. The data are fit to Equation \ref{bystander} to get an estimate of the energy difference between molecular ground state and the bystander state, and it is found to be given by $\varepsilon_b = 73.4 (\pm 1.1) meV$.}
\label{fig:bystanderstate}
\end{figure}
The peak at higher temperature is not related to the decay product, and is most likely related to a different form of the DO11 molecule, such as the DO11 molecule in an excited vibronic state in the electronic ground state manifold, an isomer, a tautomer, or twisted intramolecular charge transfer (TICT) state.\cite{westf12.01}  The important point is that whatever the species, it is not a decay product but does not contribute to ASE.  Given that this product does not participate in the photodegradation process, we refer to it as a bystander state of the undamaged molecule.  If the energy difference between the undamaged molecule and the bystander state is $\varepsilon_b$, the temperature dependence of the undamaged population is given by,
\begin{equation}\label{bystander}
N'(T) = \frac {N} {1 + \exp \left[ - \varepsilon_b / kT \right]}
\end{equation}
Figure \ref{fig:bystanderstate} shows a fit to the temperature dependence of the peak of the absorption spectrum shown in Figure \ref{fig:abstemp} from which we get $\varepsilon_b = 73.4 (\pm 1.1) meV$.  Note that at room temperature, the bystander population is about 5\%.

Generalizing Equation \ref{InteractingRates} to include the bystander state is somewhat complicated by the details of the damage mechanism.  For example, does the pump beam damage both the DO11 molecule in its undamaged and bystander state?  Is the damaged state also characterized by two species?  If the damage process removes molecules from the undamaged population, thermalization will result in an increase in the undamaged population from the reservoir of molecules in the bystander state.  Similarly, if a bystander state is associated with the damaged species, thermalization will result in a combination of both.  Combinations and permutations of these processes can lead to even more complex behavior.

Because the population of the bystander state is small relative to the undamaged DO11 population, we will generalize Equation \ref{InteractingRates} in the most straightforward way by simply replacing $N$ with $N'$, thus excluding the bystander state from consideration, or
\begin{multline}\label{eq:recoverN6}
n[t; N'(T),I] \\= \frac { \left( N' - \alpha I / \beta \right)n_0} {n_0+(N'-n_0 -\left(\alpha I / \beta  \right)) \exp \left [ - \left(N' \beta - \alpha I \right) t\right] } .
\end{multline}
By replacing $N$ with $N'$, we are not reducing the domain size but the effective population that participates in decay and recovery of ASE.  $n_0$ continues to refer to the initial undamaged population, but excludes DO11 molecules in the bystander state.  The theory can be easily generalized to account for complexities associated with individual systems.

The theory embodied in Equation \ref{eq:recoverN6} is general in the sense that it applies to a broader set of systems than those described when motivating the derivation.  For example, it has been suggested that the decay product might be a molecule with an expelled electron that is trapped in the polymer matrix,\cite{desau09.01} an aggregate formed between two molecules, or between molecules that are bound upon the exchange of a proton.\cite{embay08.01} All these mechanisms can be equally represented by our model.

\subsection{Distribution of Domains}

The recovery process is based on groupings of molecules into generic domains and the dynamics were shown to depend on domain size.  The domain size will be governed by the competition between attractive forces and thermal disordering, leading to a distribution of domain sizes.  Examples of systems with a distribution of domain sizes include ferromagnets,\cite{sakar05.01} micellized surfactant solutions,\cite{gelba96.01} liquid crystals,\cite{colli10.01} and of course dye solutions. In each case, the equilibrium  domain size distribution is associated with a minimum of the free energy. The domain size distribution can be derived in several ways.\cite{cates90.01,duque97.01,mckit10.01} The most common and simple method is minimizing the Helmholtz free energy using a grand canonical partition function.  We use this approach because our system shares with these others the interaction between entities (in our case molecules, mediated by the polymer) that form a domain and thermal interaction that limit domain size.

The partition function, $z_N$, for a domain with $N$ molecules is given by,
\begin{equation}\label{pfunction}
    z_N= \exp{\left[\frac{\lambda (N-1)}{kT}\right]}=\exp{[\gamma(N-1)]}
\end{equation}
where $\lambda$ is the free energy of a single molecule outside of a domain relative to a molecule within a domain such that the energy associated with a domain with $N$ molecules is, $E_N=-\lambda(N-1)$, $k$ is the Boltzmann constant, $T$ is the absolute temperature, and $\gamma=\lambda/kT$. When $\lambda$ is positive, energy is released when the molecule is added to a domain.  The global partition function of a collection of domains is then given by,
\begin{equation}\label{gpfunction}
    Z=\prod_N\frac{z_N^{\Omega_N}}{\Omega_N!}
\end{equation}
where $\Omega_N$ is the number of domains with $N$ molecules. If our system has unit volume, $\Omega_N$ is the number density of domains of size $N$.

The Helmholtz free energy, $F$, can be obtained from the partition function and simplified using Stirling's approximation,
\begin{align}
  \nonumber F &= -kT\ln{Z}\\
  \nonumber &= -kT\sum_N [\Omega_N\ln{z_N}-\ln{(\Omega_N!)}]\\
   &\approx kT\sum_N \Omega_N \left(  \ln \frac {\Omega_N} {z_N}  -1 \right) .
\end{align}
The chemical potential of a domain of size $N$ is therefore,
\begin{equation}\label{chempotential}
    \mu_N=\frac{\partial F}{\partial \Omega_N}=kT\ln{\frac{\Omega_N}{z_N}}.
\end{equation}

At equilibrium, the chemical potential of each molecule, whether it is a single molecule or part of a domain of any size, is the same. Equating $\mu_N/N$ and $\mu_1$, one obtains a relationship between the number density of domains of size $N$ and the number density of single molecules as,
\begin{align}\label{domainno}
    \Omega_N=e^{\gamma (N-1)}\Omega_1^N=\frac 1 {e^{\gamma}} \left[e^{\gamma}\Omega_1 \right]^N .
\end{align}

The total number of molecules in the system is given by,
\begin{align}\label{consmass}
\nonumber    \rho&=\sum_{N=1}^{\infty}N\Omega_N\\
\nonumber    &=\sum_{N=1}^{\infty}N e^{\gamma (N-1)}\Omega_1^N\\
    & =\frac{\Omega_1}{(1-e^{\gamma}\Omega_1)^2},
\end{align}
where $\rho$ can be interpreted as the average number density of molecules for fixed total volume, which is also simply related to the concentration of dye in the system.

After some rearrangement of Equation \ref{consmass}, we get
\begin{equation}\label{volfracx1}
    \Omega_1=\frac{(1+2\rho e^{\gamma})-\sqrt{1+4\rho e^{\gamma}}}{2\rho e^{2\gamma}}.
    \end{equation}
Substituting Equation \ref{volfracx1} into Equation \ref{domainno}, we can write the number density of domains of size $N$, $\Omega(N)$ as a function of number density, $\rho$,
\begin{align}\label{domainno2}
 \nonumber  \Omega(N)&=z^{(N-1)}\left[\frac{(1+2\rho z)-\sqrt{1+4\rho z}}{2\rho z^2}\right]^N \\
   &= \frac 1 z \left[\frac{(1+2\rho z)-\sqrt{1+4\rho z}}{2\rho z}\right]^N,
\end{align}
where $z=\exp\left(\frac{\lambda}{kT}\right)$.



\begin{figure}
\includegraphics{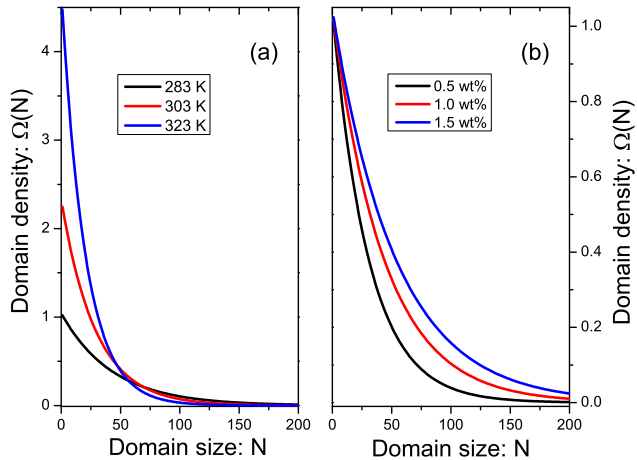}
\caption{Number density of domains $\Omega(N; \rho,T)$ as a function of domain size $N$ at several temperatures (a) and concentrations (b).}
\label{fig:domaindensity}
\end{figure}

In the model, the density, $\rho$, and temperature, $T$, are experimentally controllable, with the free energy of a single molecule outside of a domain relative to a molecule within a domain, $\lambda$, the only free parameter. Figure \ref{fig:domaindensity} shows a plot of the simulated distribution of  $\Omega(N; \rho,T)$ at several concentrations and temperatures with the other parameters fixed.  At higher temperature, the average domain size decreases as the thermal energy breaks the domains apart.

\subsection{The Full Integrated Model}

The theory of correlated molecules that describes the recovery process of a domain is governed by Equation \ref{eq:recoverN6} while the thermodynamic model describes the distribution of domain size as governed by Equation \ref{domainno2}. The effects of correlations and the statistical model for domain size can be combined through an ensemble average to predict the number density of the undamaged population as a function of time, temperature, light exposure, initial non-equilibrium undamaged population, and concentration of chromophores.

In defining the domain size in the thermodynamic model, $N$ refers to the total number of molecules in a domain, including the DO11 molecule, the bystander species and the degraded species.  However, the recovery process is hypothesized to be governed by interactions between DO11 molecules with the bystander state removed, so the bystander species must be excluded.  The mean number of undamaged molecules in a domain, $ \overline{n}$, is thus given by the ensemble average,
\begin{align}\label{recnwithT2}
 \overline{n}(t;\rho ,T,I,n_0)= \sum_{N=1}^{\infty}n(t;N',I)\Omega(N;\rho, T)\\\approx\int_1^{\infty} n(t)\Omega(N)dN \label{IntegrateDistribution}\\ \equiv \int_1^{\infty} \eta(N, I, t)dN,\label{finaln}
\end{align}
where Equations \ref{eq:recoverN6} and \ref{domainno2} are used to evaluate the integral in Equation \ref{IntegrateDistribution}, and $N'$ is given by Equation \ref{bystander}.

\begin{figure}
\includegraphics{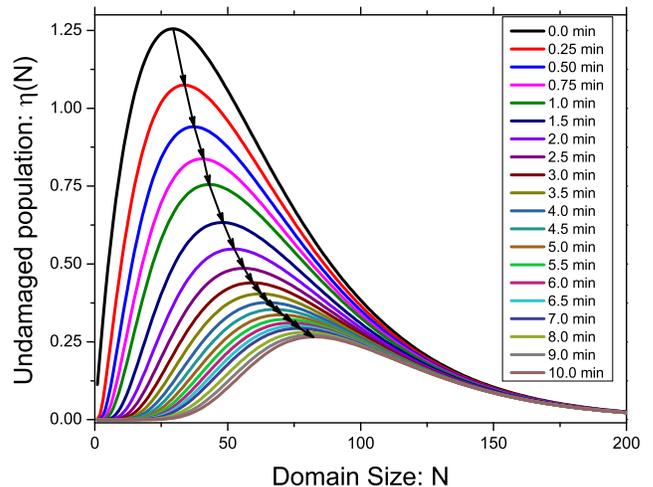}
\caption{Evolution of $\eta(N)$ as a function of time during continuous pumping.}
\label{fig:etadecay}
\end{figure}

\begin{figure}
\includegraphics{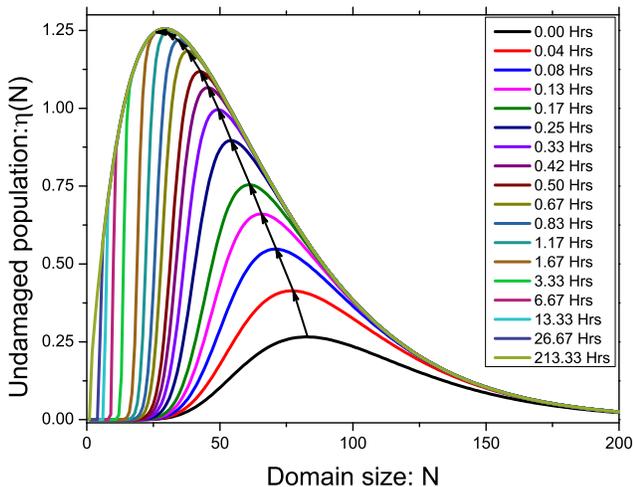}
\caption{Evolution of $\eta(N)$ as a function of time during recovery after pumping for 10 minutes.}
\label{fig:etaRecover}
\end{figure}

The distribution function $\eta(N,I,t)$ represents the population of the undamaged species and is thus proportional to the contribution to optical absorbance of the undamaged species from each group of domains of size $N$. The total ASE emission from domain size $N$ is also related to $\eta(N,I,t)$. $\eta(N) \Delta N$ is the fraction of the undamaged population living in domains of size between $N$ and $N + \Delta N$.

Figure \ref{fig:etadecay} shows simulations of the evolution of $\eta(N,I,t)$ over time as the material is pumped. Area under each curve represents the ensemble average $\overline{n}(t)$. Since molecules in smaller domains degrade at a higher rate than molecules in larger domains, the peak in the distribution shifts to the right upon photodegradation, as shown by the dashed curve.  At long times, the distribution function converges to an equilibrium shape as the decay and recovery rates balance each other in each domain.  When the pump is turned off, the damaged species in the larger domains will recover more quickly.  As the larger domains recover, the smaller domains follow, resulting in the peak position shifting back to the left, as shown in Figure \ref{fig:etaRecover}.  The peak position as a function of domain size during a run of decay and recovery thus traces a hysteresis loop.

\begin{figure}
\includegraphics{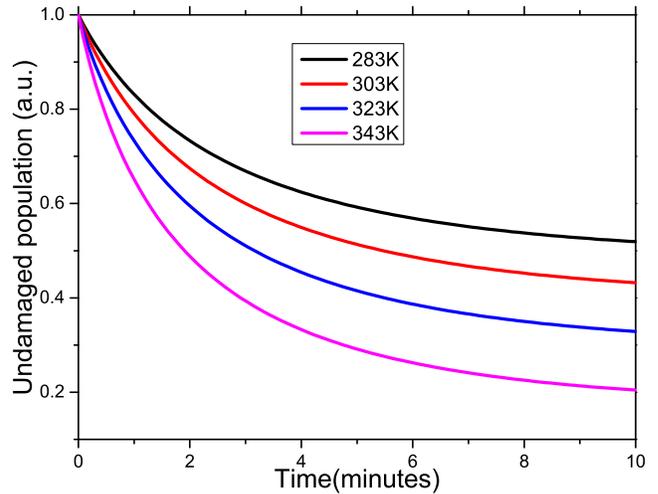}
\caption{Simulated undamaged population decay as a function of time at several temperatures at particular dye concentration and pump intensity.}
\label{fig:ASEtempSim}
\end{figure}

From Equation \ref{finaln}, we can predict the behavior of the decay of dye-doped polymer as a function of dye concentration and temperature.  Figure \ref{fig:ASEtempSim} shows a series of simulated curves of the time dependence of ASE as predicted for various temperatures.

\section{Testing the Theory}

Experimental details are described elsewhere\cite{ramin11.01}. It is important to note that due to spatial variations from spot to spot, it is difficult to get reproducible data even from a single sample.  Typical point to point sample variations can yield  as much as 30\% variation in the signal, with larger variations from sample to sample.  As such, most measurements require multiple runs to be averaged to decrease the effects of such variability.

In determining the change of optical absorbance, the white light source - which focuses to a circular area, must be overlapped with the region that generates ASE.  Since the pump light must be focused to a thin line to optimize ASE signal, the pump light and the wider white probe beam can never fully overlap.  It is also difficult to assure that the overlap region is optimized, so, most of the probe might be missing the damaged area.  To make matters worse, the pump beam can have hot spots that evolve over long data runs - which can take up to several days.  Thus, some areas along the pump line may be fully damaged while others are close to pristine.  This makes it difficult to definitively determine the change in the absorption spectrum for a damaged area.  To overcome these issues requires painstaking adjustments and multiple runs in different spots on one sample as well as runs on different samples prepared in the same way.

\begin{figure}
\includegraphics{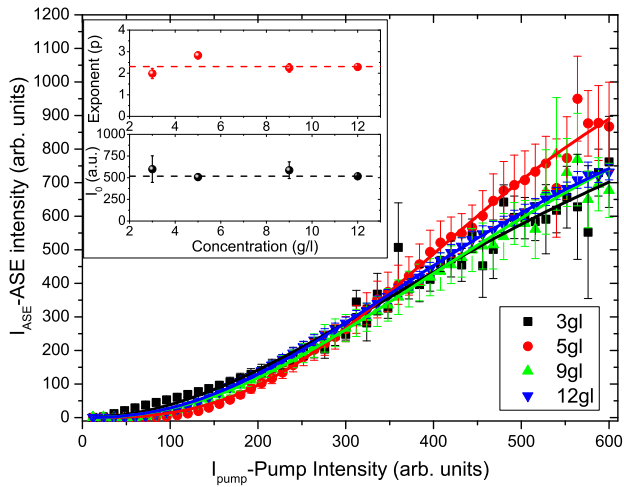}
\caption{ASE intensity plotted as a function of the pump intensity for samples of several concentrations. The inset shows the fit parameters $I_0$ and exponent $p$ determined from the data at each concentration.}
\label{fig:ASEvsPump}
\end{figure}

Empirically we find that the ASE intensity satisfies,
\begin{equation}\label{ASEexpression}
    I_{ASE}=\frac{(c/c_0)^q}{1+(I_0/I_{pump})^p},
\end{equation}
and that the undamaged population is governed by the equation
\begin{equation}\label{conc-VS-ASE}
\overline{n} \propto [I_{ASE}]^{1/2.6}
\end{equation}
for a particular pump intensity, where where $I_0$, $p$, $c_0$ and $q$ are constants and where $c$ is the dopant concentration and $I_{pump}$ is the intensity of the pump. Figure \ref{fig:ASEvsPump} shows the ASE intensity as a function of the pump intensity for several concentrations of DO11 dye as well as the fit to Equation \ref{ASEexpression}.  The curves are normalized for the figure to make them overlap to show that they all have the same shape.  The shape of the curve for the sample of $5 g/l$ concentration is somewhat anomalous.

\begin{figure}
\includegraphics{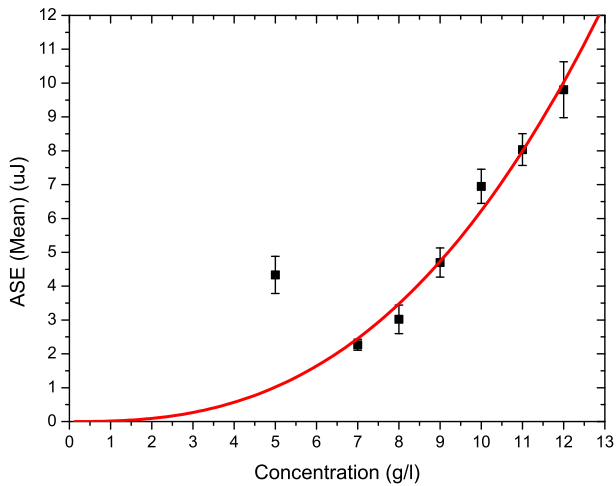}
\caption{Averaged ASE intensity at several dye concentrations. The data is fit to the Equation $I_{ASE} \propto (c/c_0)^q$ with $c_0 = 5.0(\pm0.6)~g/l$ and $q = 2.6 \pm 0.5$.}
\label{fig:ASEvsConcentration}
\end{figure}
Figure \ref{fig:ASEvsConcentration} shows a plot of the ASE intensity as a function of dye concentration and a fit to the function $I_{ASE} \propto (c/c_0)^q$, which yields $c_0 = 5.0(\pm0.6)~g/l$ and $q = 2.6 \pm 0.5$.  Thus, the ASE intensity can be used as a measure of the concentration of undamaged species through this relationship.

As a test of the theory's ability to predict the dependence of the decay of ASE intensity as a function of time, samples of several concentrations are characterized using ASE at fixed temperature and pump intensity.  The ASE intensity is converted to population of undamaged DO11 molecules using Equation \ref{conc-VS-ASE}.  Initially the data from one concentration is fit to the model given by Equation \ref{finaln} by independently varying all the parameters ($\alpha, \beta, \rho$, and $\lambda$). Subsequently, the rest of the concentrations are fit to the model keeping constant the values obtained for $\alpha$, $\beta$, and $\lambda$ above and allowing only $\rho$, the number density of DO11 molecules, to vary. A value of $\rho$ is determined for each of the concentrations.

\begin{figure}
\includegraphics{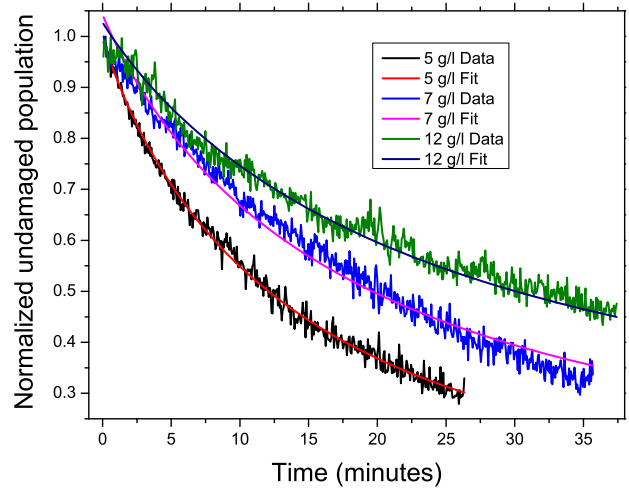}
\caption{Normalized undamaged population ($\overline{n}$) obtained from the ASE intensity is fitted to the correlation model to get the values of parameter $\rho$.}
\label{fig:ConcDecayFits}
\end{figure}

Figure \ref{fig:ConcDecayFits} shows representative data and fits to the theory for three concentrations.  Six different concentrations are tested, and multiple runs at each concentration yield multiple values of the fit parameter $\rho$, which are averaged.  The experimental uncertainty is determined from the spread in the data.  Table \ref{tab:results} summarizes the values of the three parameters obtained from the fits. These same values of the three parameters are applied to the the full set of data in this paper.  $\alpha$, $\beta$ and $\lambda$ are properties of the material, and therefore should be constant for a given polymer and dye combination, though variations in the polymer due to materials processing may result in slight differences in these constants.
\begin{figure}
\includegraphics{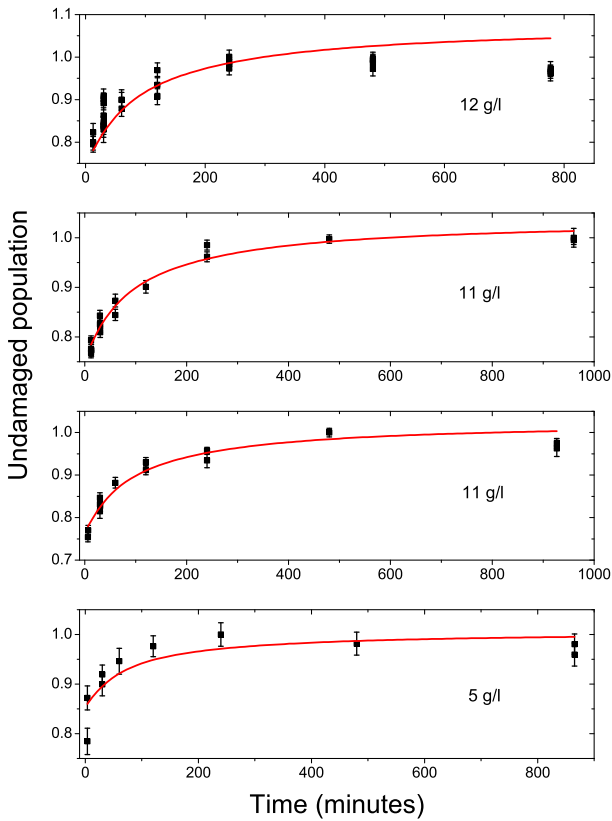}
\caption{Recovery of undamaged population ($\overline{n}$) and a fit to the correlated chromophore model using the parameters obtained from fits to the decay of population during pumping.}
\label{fig:ConcRecoverFits}
\end{figure}

\begin{center}
\begin{table}
\caption{Parameters determined for DO11 in PMMA using an average pump intensity of $I_p = 0.202 W/cm^2$.\label{tab:results}}
  \begin{tabular}{c  c  c }
  \hline
    $\alpha (min^{-1}W^{-1}cm^2)$& $\beta (10^{-4}min^{-1}) $ & $ \lambda (eV)$  \\
    Decay Rate & Recovery Rate & Free energy/molecule \\ \hline\hline
    $7.09 (\pm 0.13)$ & $3.22 (\pm 0.26)$ & $0.29 (\pm 0.01)$ \\ \hline
  \end{tabular}
\end{table}
\end{center}

Using the parameters determined from the decay data, the recovery data is predicted without the use of adjustable parameters.  Figure \ref{fig:ConcRecoverFits} shows the predicted population as a function of time (curves) during recovery and the measured population of undamaged DO11 molecules as determined from the ASE intensity.  We note that different samples were used to study recovery.  Even so, the theory's prediction agrees with the data within experiential uncertainties for a range of concentrations.

\begin{figure}
\includegraphics{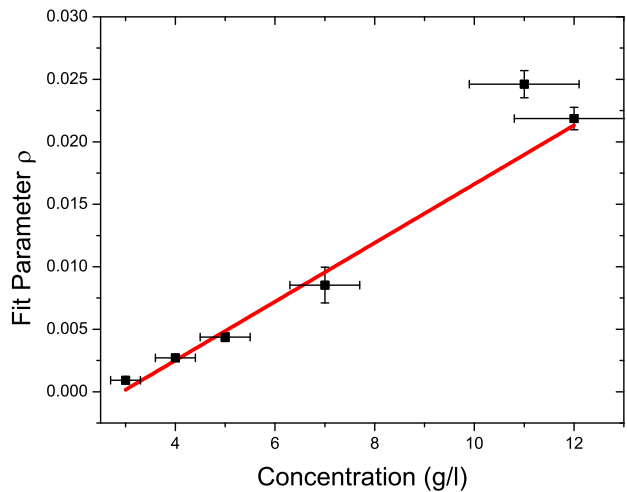}
\caption{Fit parameter $\rho$ from decay fits as a function of dye concentration as determined during sample preparation.}
\label{fig:ConcLinearFit}
\end{figure}
According to the model, $\rho$ is proportional to the concentration of dyes in the polymer. As such, we expect a linear relation between $\rho$ and concentration. Figure \ref{fig:ConcLinearFit} shows a linear fit to a plot of $\rho$ as a function of the concentration of a series of samples as determined from the amount of dye added to the polymer during its preparation.  The linear fit is consistent with the data.

\section{Conclusions}

We have presented a simple model for photodegradation and recovery that depends on three parameters: the intensity-dependent decay rate, the recovery rate, and the free energy of a single molecule outside of a domain relative to a molecule within a domain.  This model accounts for all observations of ASE and absorption spectroscopy of DO11 dye in PMMA as a function of time, pump intensity and concentration during decay and recovery with one set of these three parameters.  Furthermore, the theory predicts the behavior as a function of temperature, which we are in the process of testing with experiment in which preliminary data is consistent with predictions.  A more exhaustive experimental study will be presented in the future.

We find that three parameters fully characterize a composite material made from a particular chromophore and host polymer.  As such, $\alpha$, $\beta$ and $\lambda$  -- which should be calculable from first principles based on the underlying mechanisms -- hold the key in providing a connection between experiment and more fundamental theoretical considerations.  $\alpha$ is related to the damage cross-section of the molecule, which is empirically found to decrease with increased domain size.  $\beta$, on the other hand, characterizes the self-healing process, with $n \beta$ a measure of the collective strength of undamaged molecules to heal.  The parameter $\lambda$, on the other hand, determines the distribution of domain sizes, which are affected by temperature and the concentration of the molecules in the sample when it is prepared.

The theory is easily generalizable to account for other observations.  For example, all studies to date where self healing is observed conclude that the degradation rate is proportional to the intensity, implying a linear mechanism of damage is responsible.  For a nonlinear damage process, Equation \ref{NonInteractingRates} can be generalized by adding a nonlinear function of the intensity.  If an irreversible damage mechanisms acts along with a reversible one, an additional species can be added to the model.

The mechanism responsible for self-healing appears to be in the interactions between molecules that are somehow mediated by the polymer.  Indirect evidence suggests that damage is associated with charge ejection from pristine molecules and the creation of a trapped charge\cite{zimme94.01} density in the polymer.\cite{desau09.01} We propose the hypothesis that in liquid solution, molecules that are damaged may break apart into fragments (perhaps charged) that by virtue of mixing in the liquid state takes them too far apart, on average, to recombine over a reasonable time frame.  Once the molecular ions are neutralized by charges in the liquid through collisions and mass transport, reconstitution of the original molecule from the fragments is energetically unfavorable, and the process is thus irreversible.  In this hypothesized mechanism, the ionic fragments separate, but because of the polymer, remain in closer proximity to each other. Electrostatic attraction between the fragments drives recombination.  If molecules are associated with others in a domain, the larger fragments remain relatively stationary compared with the lighter ones.  Thus, as higher numbers of fragments are produced, attractive forces to the domain increases, enhancing self-healing, as experimentally observed.  Dielectric screening due to the polymer may act to lengthen the healing time.

Another possibility, though less likely, is that self healing is a process that is a totally new phenomena analogous to Bose-Einstein condensation, which favors relaxation into a Fock State in which all particles are in the same single-particle state.  In this picture, the molecules are correlated due to positive exchange statistics of some sort, where a domain of undamaged molecules induces healing in proportion to their population.

The source of correlations between molecules is not clear; but, there are many possibilities.  It has been suggested that the association may be in the form of aggregation,\cite{kuzyk06.06,embay08.01} which can originate in either electrostatic interactions, hydrogen bonding, or the formation of nanocrystalites.  Alternatively, phonons in the polymer chains - which behave as bosons, may result in correlations.  Or, the mechanism may be an altogether new phenomena.   Any process in which an aggregate is of lower energy, or is more probable due to exchange statistics, is a viable candidate. Independent measurements are needed to sort them out.

There are many experiments that can be brought to bare on the problem of mechanisms.  The role of the polymer and its interaction with a guest molecule can be tested by varying the properties of each.  For example, the fact that self healing is not observed in liquid monomer but in polymer suggests that there exists a critical level of polymerization that leads to self healing.  As such, one could measure the healing rate as a function of polymerization, is situ -- while a dye-doped monomer is in the process of polymerizing -- and correlate molecular weight with healing to determine if each domain is composed of molecules that are associated with a single polymer chain.  Alternatively, the material can be heated through the polymer glass transition to determine if increased mobility of the polymer chains interferes with the healing phenomena.  In addition, the polymer host and guest molecule can be changed to determine the importance of structural and chemical properties to healing.

Optical characterization, including various types of imaging, ASE, absorption spectroscopy, and fluorescence can be applied in tandem with other experiments such as photoconductivity to test a given hypothesis.  For example, an observation of photoconductivity in coincidence with changes in population of damaged species as probed optically would support the hypothesis that charge ion generation plays a role.  Suppression of healing in the presence of a strong electric field would support the mechanism of charged molecular fragments, as would thermally stimulated discharge measurements.  Finally, neutron scattering experiments could be used to directly probe correlations and aggregation.

In summary, the model that we present here predicts all of the observations in terms of three parameters and is based on the hypothesis that self healing originates from a collective phenomena in which pristine molecules induce healing in a damaged molecule in proportion to domain size.  Similarly, the decay rate decreases with domain size.  Future planned experiments will be used to zero in on the mechanisms.

{\bf Acknowledgements:} We thank the Air Force (Grant No:~FA9550-10-1-0286) for their generous support and Dr.~Fred Gittes for insightful discussions.


\end{document}